# CLIC MDI OVERVIEW

Lau Gatignon[1] on behalf of the CLIC MDI Working Group

1 – CERN – EN Department
1211 Geneva 23 – Switzerland

This paper gives an introduction to the layout of the CLIC Machine Detector Interface as it has been defined for the CLIC Conceptual Design Report. We concentrate on the specific case of the CLIC_SiD detector, although the push-pull concept for two detectors has been included in the design. Some recent work and developments are described as well. However, for the details we refer to the detailed technical talks at this conference.

## 1 Introduction

This paper gives an introduction to the layout of the CLIC Machine Detector Interface as it has been defined for the CLIC Conceptual Design Report. We concentrate on the specific case of the CLIC_SiD detector, although the push-pull concept for two detectors has been included in the design. Some recent work and developments are described as well. However, for the details we refer to the detailed technical talks at this conference.

The CLIC Machine Detector Interface is driven by a number of key parameters of the CLIC machine. In particular the very small vertical spot size of 1 nm RMS at the interaction point leads to some very challenging requirements on the final focus quadrupole QD0 and it alignment and stabilization. To reach the maximum luminosity, the QD0 is located at a distance $L^* = 3.5$ m from the interaction point and therefore inside the detectors.

The CLIC project foresees one interaction point (IP), but two detectors. Therefore the detectors will be mounted on movable platforms that allow to move the detectors onto the IP or away into a service cavern. It is assumed that interventions that require opening of the detector will in any case not be done at the IP but in a service cavern. This is illustrated in Figure 1. Please note that the outer dimensions of the CLIC detectors are very similar.

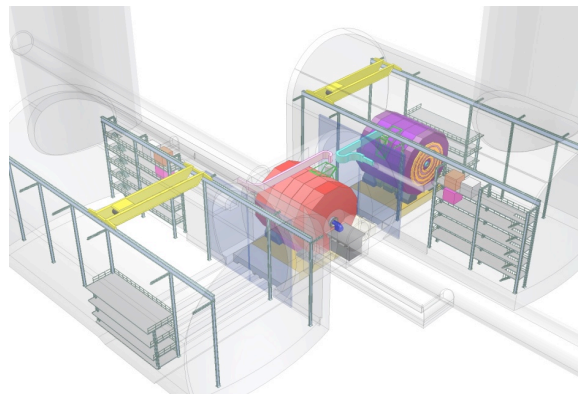

Figure 1: The cavern layout

The Machine Detector interface includes not only the final focus quadrupoles and their support tubes, but also anti-solenoids to protect the QD0 magnets against the main solenoid field and to minimize the main solenoid impact on the beam, QD0 pre-alignment and stabilization equipment, beam position monitors and kickers for an intra-pulse feedback system, beam and luminosity monitoring calorimeters and passages for the spent beam towards the final beam dumps. The vacuum system must be able to provide the required vacuum performance and segmented to allow push-pull operation. A schematic drawing of the proposed layout is shown in Figure 2.

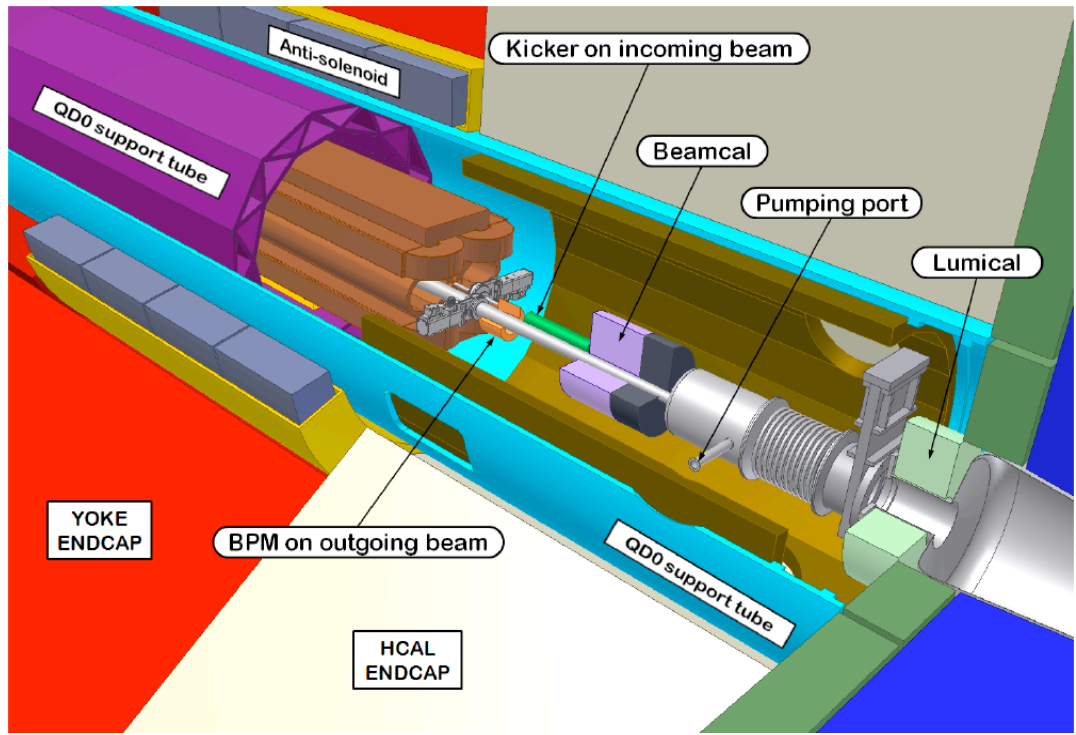

Figure 2: A schematic view of the CLIC Machine Detector Interface

## 2  The Final Focus Quadrupole Magnet QD0

The QD0 magnet has an aperture radius of 3.83 mm, a length of 2.73 m and a nominal gradient of 575 T/m. This corresponds to a peak field of 2.2 T. The outer radius may not exceed 35 mm in certain directions to allow the passage of the spent beam. The field must be tunable over at least 10% towards lower values, but a larger tuning range would be an advantage.

On top of these challenging parameters, the design must be compatible with stringent stabilization requirements, namely to 0.15 nm RMS for frequencies above 4 Hz. Based on these specifications, we have opted for a hybrid technology. The magnet yokes are made of Permendur, an alloy with a high magnetic saturation. Coils provide the variable magnetic field and permanent magnets boost the performance ensuring that the magnetic field leakage from the Permendur poles due to saturation is minimized. For the permanent magnets the baseline is $Sm_2Co_{17}$ as magnetic material. The design is calculated to provide 531 T/m gradient, which can be increased to 590 T/m with $Nd_2Fe_{14}B$, which is however less radiation hard. To reduce vibrations, the coils are supported independently from the magnet and the current density is kept very low to allow zero or very low cooling water speeds in the indirect cooling system. A schematic drawing of the QD0 magnet is shown in Figure 3 and of the support tubes in Figure 4.

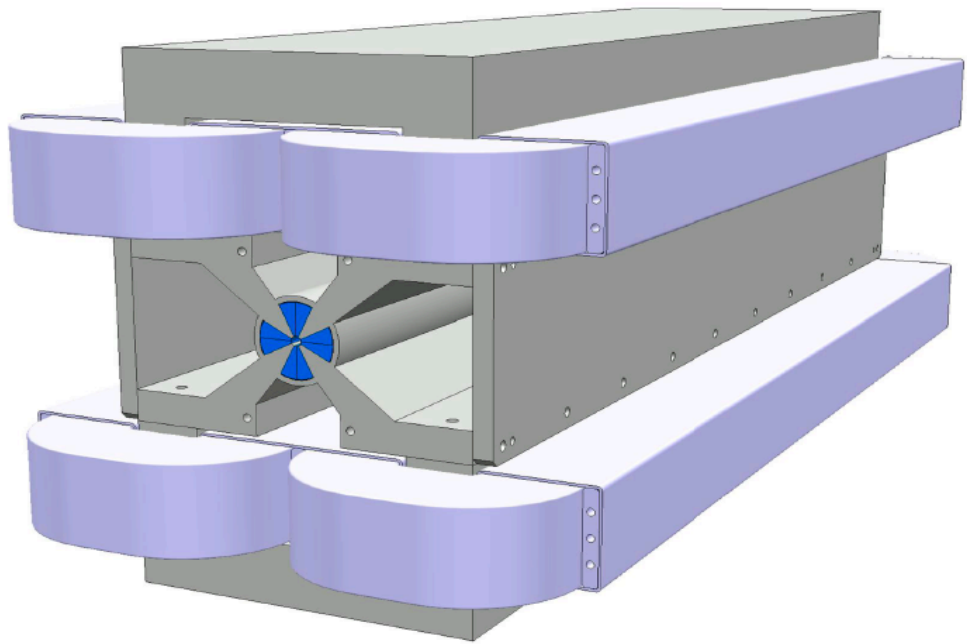

Figure 3: A sketch of the QD0 magnet

A short (10 cm) prototype is in preparation at CERN. A picture of the Permendur yokes and permanent magnets is shown in Figure 5. First field measurements without the coils have been performed. The coils are being manufactured and first tests with the coils powered are expected before the end of 2011.

Details of the QD0 design and the prototype construction and tests are presented in [1].

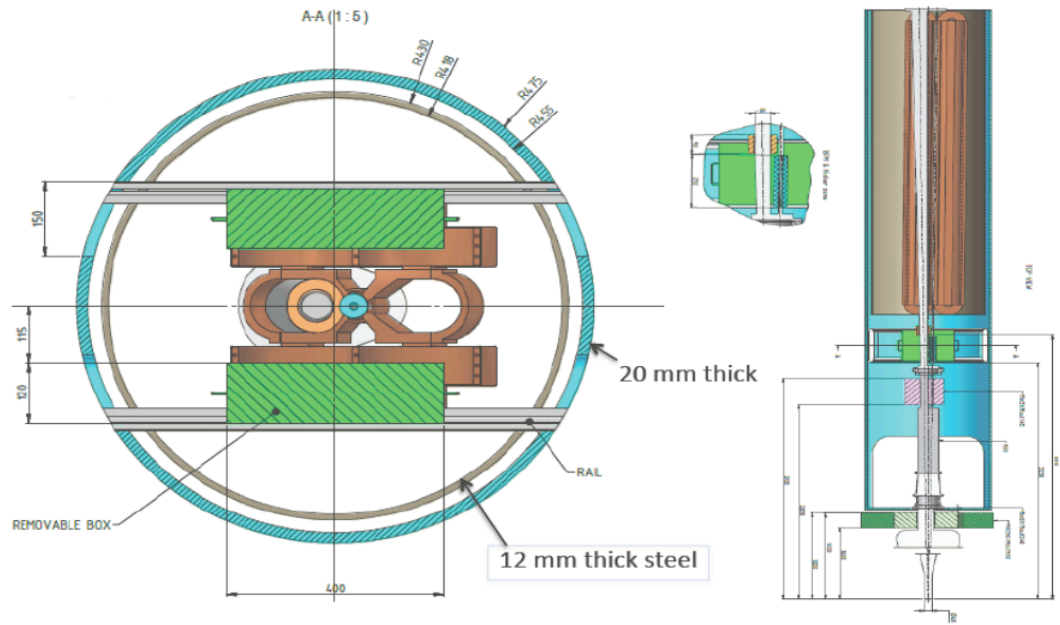

Figure 4: A sketch of the double QD0 support tube

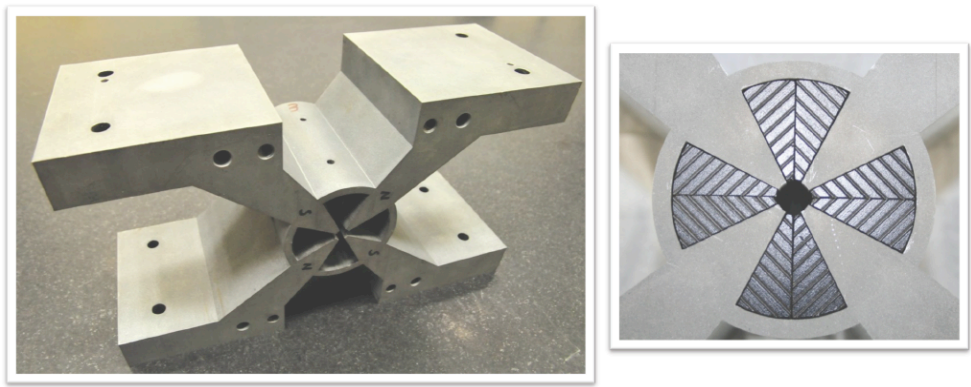

Figure 5: A photograph of the QD0 short prototype without its coils

## 3 The anti-solenoid

The permanent magnets inside the QD0 magnet must be protected against the external fields of the main experiment solenoid. Also it is important to shield the incoming beam against the effect of the main solenoid, because with the small crossing angle (20 mrad) the field would reduce the luminosity of the CLIC machine. In the CDR it was shown that a first design of the anti-solenoid allowed keeping the luminosity loss well within the foreseen luminosity budget. However, engineering aspects were not addressed yet. Recently 3D magnetic simulations of the entire MDI region have been performed. These allow to properly estimate the mechanical forces provoked by the coupling of the various fields (main solenoid, anti-solenoid and QD0) and to optimize the geometry and field configuration accordingly. This work is in progress and was presented in [2], together with results on the field compensation and the forces involved. A sketch of the present layout is shown in Figure 6.

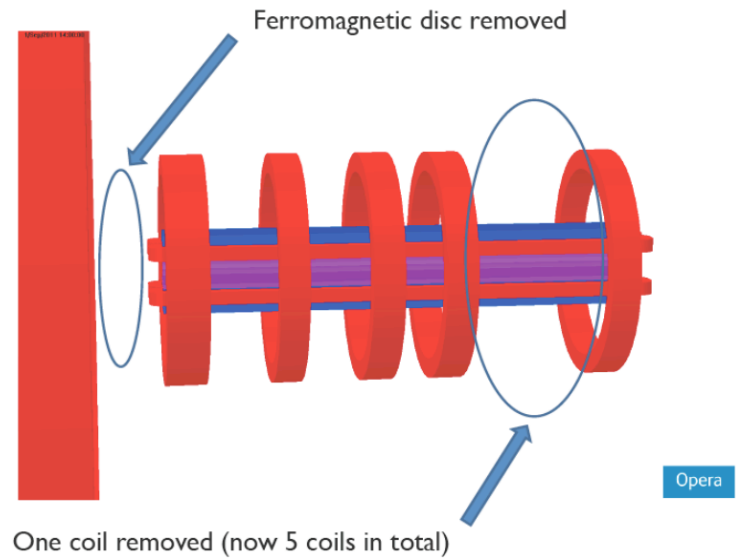

Figure 6: A sketch of the latest anti-solenoid design

## 4  QD0 stabilization

The QD0 position must be stabilized mechanically to within 0.15 nm RMS for frequencies above 4 Hz. Relative movements will be detected by capacitive sensors and the absolute position by a geophone in a shielded region nearby. The position measurements are fed back to a piezo actuator that adjusts the position according to the sensor output. Active stabilisation to 0.13 nm at 4 Hz has been demonstrated in the lab [3], but on a massive and voluminous support. The capacitive sensors and the actuators are mounted in a compact stabilization feet, installed below the QD0 quadrupole inside its support tube. A preliminary design of the stabilization foot is shown in Figure 7. Details of the stabilization approach and of the integration into the MDI environment are presented in [4].

The stabilization scheme requires the stabilization device and the QD0 quadrupole to be located in a relatively stable requirement. The expected ground motion in the CLIC tunnel is at the level of several nanometers RMS in the relevant frequency range. A passive isolation system, called pre-isolator, is foreseen to relieve this situation.

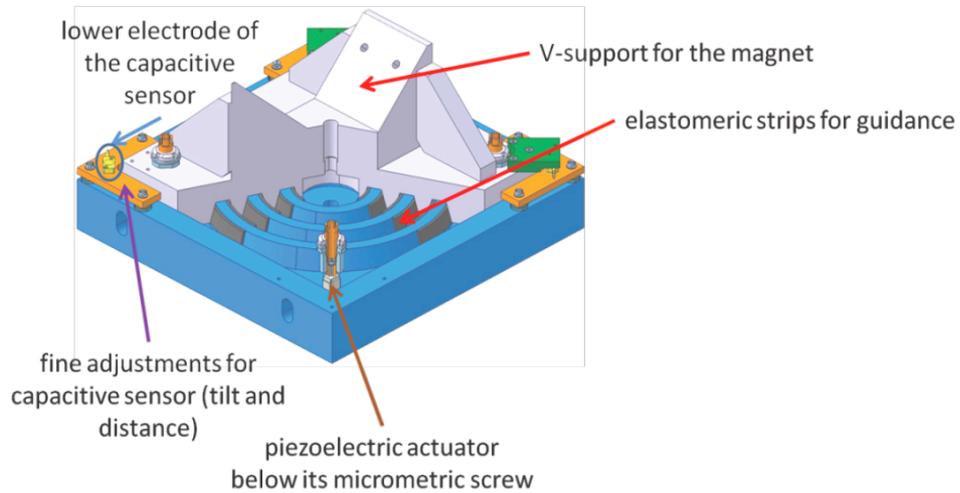

Figure 7: A preliminary design of the stabilization foot

The pre-isolator consists of a massive concrete block (~80 tons) mounted on calibrated springs with an eigen-frequency around 1 Hz. See Figure 8. According to simulations this system could reduce ground motion effects by a factor of up to 30 for the higher frequencies, however at the cost of an increased amplitude at lower frequencies (1 Hz and below). However, the effects of such low frequency vibrations can be mitigated by beam feedbacks. More details of the pre-isolator are shown in [5].

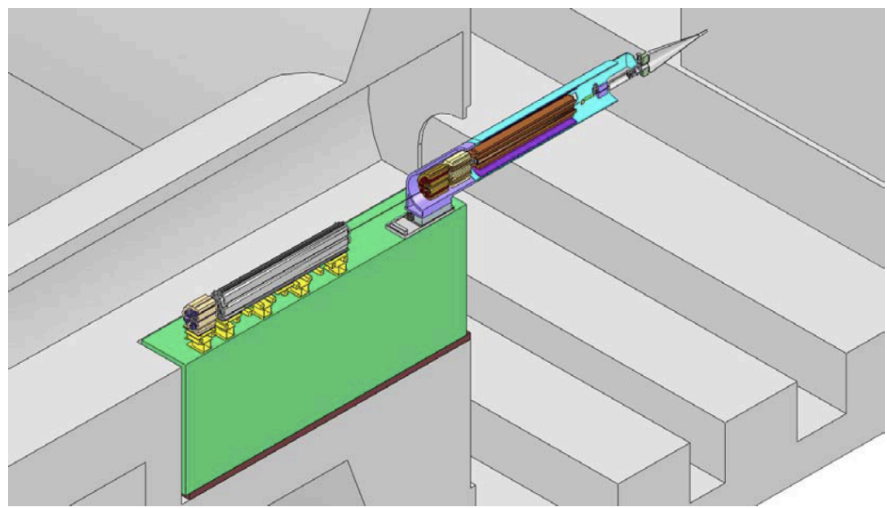

Figure 8: A schematic view of the pre-isolator

# 5 QD0 pre-alignment

The QD0 magnets have to be pre-aligned to within 10 microns with respect to each other and with respect to the other beam elements in the Beam Delivery System. The alignment with respect to the Beam Delivery System is based on stretched wires. See [6] for more details. For the relative alignment of the two QD0 magnets, a network of RASNIK systems is used. This system is described in some detail in [7]. For this system to work, light channels must be implemented through the detector. The RASNIK systems measures the positions of four highly precise reference rings, mounted on each QD0 extremity and equipped with proximity sensors. Six radial spokes of extremely well-known dimensions transport the light outward to alignment systems. Light channels pass the light from one end of the detector to the other in between the calorimeters and the main solenoid yoke. This is shown schematically in Figure 9. By combination of redundant information, the positions of the centers of the four reference rings are computed. The QD0 positions are subsequently adjusted with CAM movers with 5 degrees of freedom.

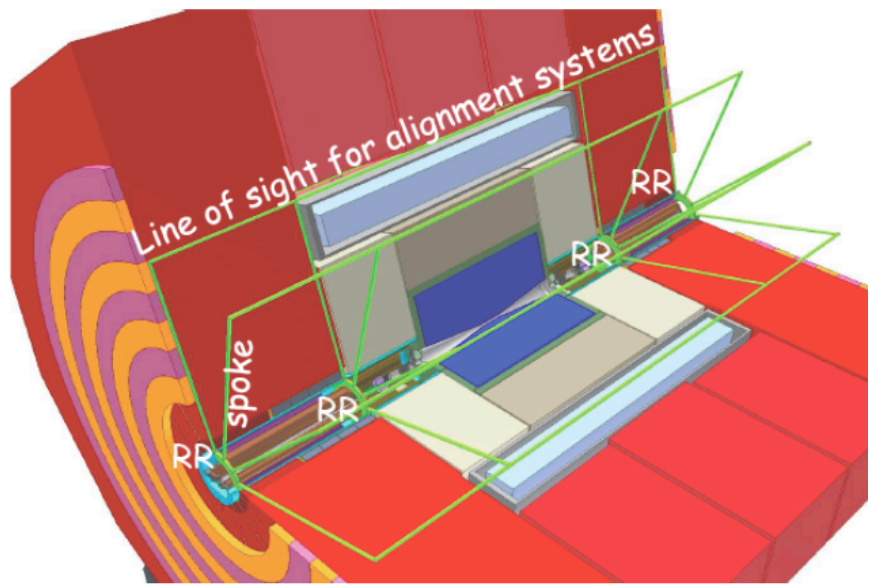

Figure 9: The light paths through the detector for the RASNIK system

# 6 IP feedback system

The beam arrives in trains of 316 bunches separated by 0.5 ns over a period of 158 ns in total. Any relative movement of the two beams over that period will lead to a loss of luminosity. A feedback system is proposed to mitigate this effect.

A beam position monitor (BPM) is located on the spent beam at a distance of ~ 3 metres from the interaction point. The position measurement is processed by fast analog electronics that within 10 ns calculates a correction via a fast kicker magnet on the opposite incoming beam. The total latency of the system can be reduced to 37 ns, dominated by travel time of the particles to and from the interaction point. With this latency it turns out to be possible to have 4 iterations within each bunch trains. The layout and performance of this IP feedback system is presented in [8]. The integration in the Machine Detector Interface is illustrated in Figure 10.

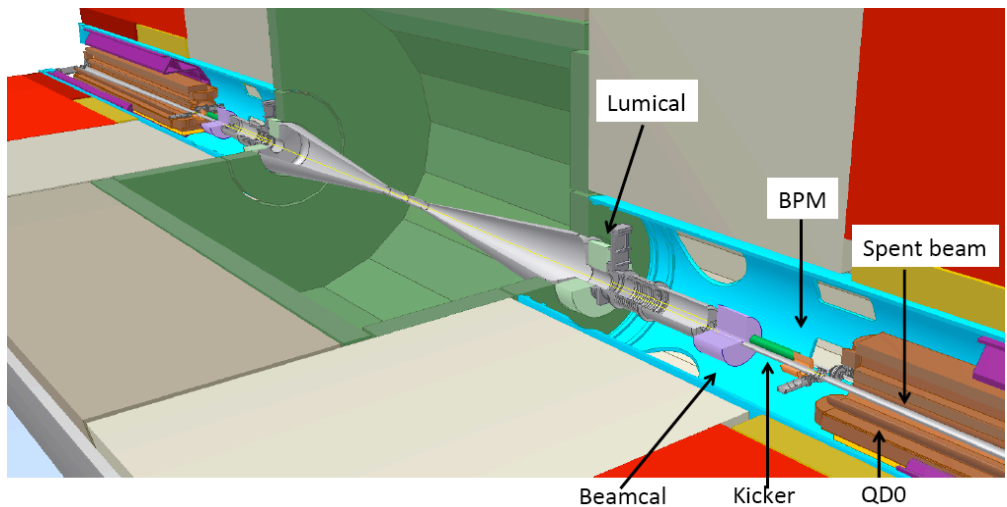

Figure 10: The integration of the BPM and kicker in the Machine Detector Interface

## 7 The detector cavern and detector movements

Thoughts have been given to a detector layout compatible with two detectors in push-pull mode. The underlying principle is that no interventions that require opening of the detector are done on the IP. The detectors are compact and have the same length. In the case of CLIC_ILD the overall length is minimized by adding compensating coils on the return yoke. This allows to contain the main solenoid field in a reduced length. Two large caverns on either side of the IP allow interventions on the detectors. There are large shafts above each of these service caverns, but not above the interaction point. Large sliding doors make the separation between the service caverns and the IP, in particular in terms of air flow. If necessary, these doors can be made thick enough to serve as radiation shielding. The schematic cavern layout is shown in Figure 11. The detectors move on large platforms. Proposals for these platforms and for the push-pull operation itself are shown in [9].

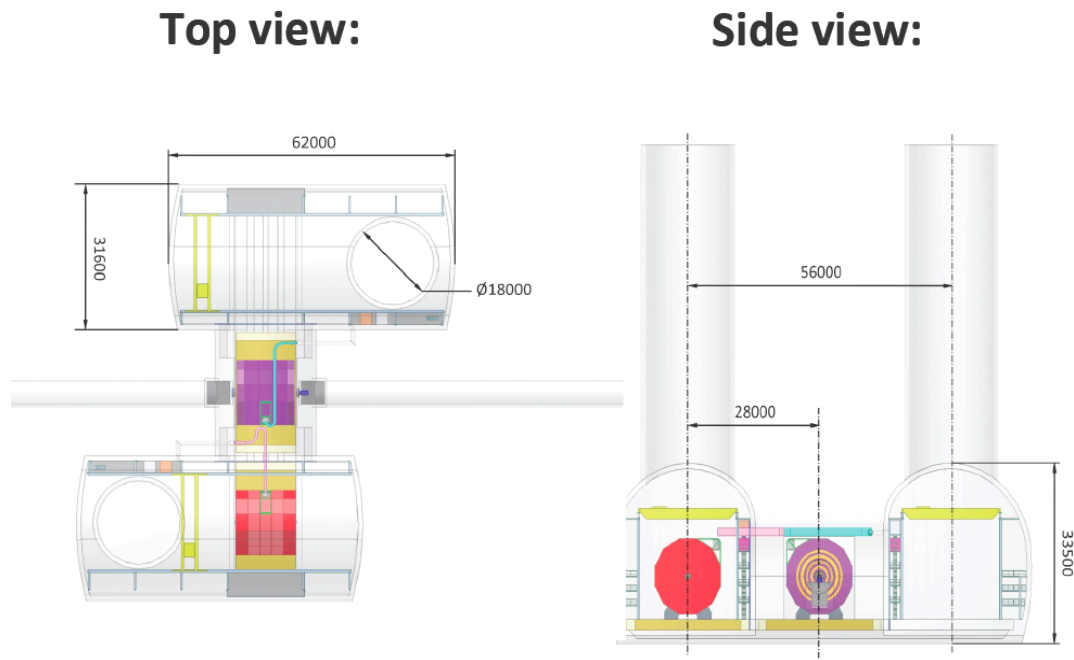

Figure 11: A schematic view of the detector cavern

## 8 The beam delivery system

The beam delivery system (BDS) is the part of the CLIC machine that transports the beam from the exit of the main linac to the interaction point. The length of each BDS is about 2.75 km. Important functions of the BDS are the beam cleaning by a collimation system, mitigation of muon backgrounds and the final focusing to provide the nanometer spot sizes.

Significant progress was reported on the tuning of the final focus system, both for a traditional final focus system [10] and for a system with the QD0 located in the tunnel at a distance $L^* = 6$ m from the interaction point [11]. The ongoing work towards reaching small spot sizes at ATF2 was shown in [12]. At the IP the beams meet at a crossing angle of 20 mrad. This angle leads to a severe loss of luminosity. This is prevented by rotating the bunches so that they collide parallel, using crab cavities. The present status of the crab cavity studies was presented in [13].

For the collimation system further studies were presented in [14]. One of the main difficulties is to make the collimators survive a full beam impact.

In normal operating conditions only a very small fraction of the beam hits the collimators. However, the high-energy electrons and positrons in the tails produce high-energy muons that may reach the detectors. Without additional measures, more than 10 muons per bunch crossing may reach the detector. In [15] it is shown that with additional muon spoilers a reduction to about 1 muon per bunch crossing is within reach. These spoilers are toroids of 55 cm outer radius and with a central hole of 1 cm radius. The field is of the order of 1.5 T.

# 9 Acknowledgements